\documentclass[onecolumn,authoryear]{els-mrw} 

\usepackage{amsmath,amssymb,amsfonts,amsthm,makeidx,graphicx}
\usepackage{txfonts}
\usepackage{helvet}
\usepackage{sidecap}
\sidecaptionvpos{figure}{m} 
\usepackage[font=normal]{subfig}
\usepackage{multirow}
\usepackage{makecell}
\usepackage[labelfont=bf]{caption}
\usepackage[colorlinks=true,citecolor=blue,urlcolor=blue]{hyperref}

\newcommand{\snia}{SN~Ia}
\newcommand{\sneia}{SNe~Ia}
\newcommand\ion[2]{#1\,\,{\sc{\romannumeral #2}}}
\newcommand{\nifs}{\ensuremath{^{56}\rm{Ni}}}
\newcommand{\nife}{\ensuremath{^{58}\rm{Ni}}}
\newcommand{\cofs}{\ensuremath{^{56}\rm{Co}}}
\newcommand{\feff}{\ensuremath{^{54}\rm{Fe}}}
\newcommand{\fefs}{\ensuremath{^{56}\rm{Fe}}}
\newcommand{\msun}{\ensuremath{\rm{M}_{\odot}}}

\newcommand{\lsun}{\ensuremath{\rm{L}_{\odot}}}
\newcommand{\rsun}{\ensuremath{\rm{R}_{\odot}}}

\newcommand{\ms}{\ensuremath{\rm{m\,s}^{-1}}}

\newcommand{\gcc}{\ensuremath{\rm{g\,cm}^{-3}}}
\newcommand{\kgcm}{\ensuremath{\rm{kg\,m}^{-3}}}

\newcommand{\js}{\ensuremath{\rm{J\,s}^{-1}}}

\newcommand{\mch}{\ensuremath{M_{\rm Ch}}}
\newcommand{\ye}{\ensuremath{Y_{\rm e}}}

\newcommand{\micron}{\ensuremath{\mu\rm{m}}}

\newcommand*\aap{A\&A}

\newcommand*\aj{AJ}

\newcommand*\apj{ApJ}
\newcommand*\apjl{ApJL}

\newcommand*\apjs{ApJS}

\newcommand*\araa{ARA\&A}

\newcommand*\mnras{MNRAS}

\newcommand*\nat{Nature}

\newcommand*\pasa{PASA}

\newcommand*\pasp{PASP}

\newcommand*\prc{Phys.~Rev.~C}

\newcommand*\prl{Phys.~Rev.~Lett.}

\begin{document}

\chapter{Type Ia supernovae}\label{chap1}

\author[1,2]{Stéphane Blondin}%

\address[1]{\orgname{Aix Marseille Univ, CNRS, CNES, LAM}, \orgaddress{Marseille, France}}
\address[2]{\orgname{European Southern Observatory}, \orgaddress{Karl-Schwarzschild-Straße 2, Garching, D-85748, Germany}}

\articletag{Chapter Article tagline: update of previous edition,, reprint..}

\maketitle

\begin{abstract}[Abstract]
Type Ia supernovae (\sneia) correspond to the thermonuclear explosion
of a carbon-oxygen white dwarf (C-O WD) star in a binary system,
triggered by the accretion of material from another star, or the
merger/collision with a secondary WD.  Their phenomenal luminosity ---
several billion times that of the sun --- has motivated their use as
cosmological distance indicators and led to the discovery of the
accelerated expansion of the universe.  \sneia\ are also the main
producers of iron and hence play a fundamental role in the chemical
evolution of galaxies.  While recent observations have confirmed the
basic theoretical picture of an exploding C-O WD star whose luminosity
is powered by the radioactive decay of \nifs, a number of
uncertainties remain concerning the nature of the binary companion and
the explosion mechanism. Several lines of evidence point towards the
existence of multiple progenitor channels in order to explain the full
range of the observed diversity. A complete physical understanding of
these energetic stellar explosions remains a long-lasting goal of
modern astrophysics.
\end{abstract}

\begin{glossary}[Glossary]

\term{Chandrasekhar mass (\mch)}: Maximum mass for a white dwarf (WD)
star; $\mch \approx 1.4$\,\msun\ for a cold, non-rotating,
non-magnetic carbon-oxygen (C-O) WD.
 
\term{Compton scattering}: In the context of Type Ia supernovae (\sneia), the scattering of
$\gamma$-rays by electrons.
   
\term{Deflagration-to-detonation transition (DDT)}: Transition of the
combustion mode from a subsonic deflagration to a supersonic
detonation.

\term{Double degenerate (DD) scenario}: Progenitor channel for
\sneia\ in which a WD merges or collides with another WD.

\term{Electron degeneracy}: Quantum-mechanical effect whereby a
confined electron gas will exert a pressure even at zero temperature.

\term{Electron fraction (\ye)}: Ratio of the number of electrons to
nucleons (protons + neutrons) in a given chemical mixture.

\term{Forbidden line}: Spectral line resulting from the emission of a
photon in an atomic transition with a very low transition probability.

\term{Homologous expansion}: Holds when the velocity $v$ of a fluid element
is directly proportional to its radius $r$: $v(r) \propto r$.

\term{Iron-group elements (IGE)}: Elements with atomic numbers in the
range $Z=21\mathrm{-}28$ (i.e., scandium to nickel).

\term{Intermediate-mass elements (IME)}: Elements with atomic numbers
in the range $Z=11\mathrm{-}20$ (i.e., sodium to calcium).

\term{Local thermodynamic equilibrium (LTE)}: State of equilibrium
between matter and radiation at the local temperature.
   
\term{Metallicity}: Abundance of all elements heavier than hydrogen and
helium.

\term{Nebular phase}: Phase in the \snia\ evolution when the ejecta are
transparent and the deposited energy is quasi-instantaneously re-emitted.

\term{Nuclear statistical equilibrium (NSE)}: State of balance in
nuclear reactions reached in explosive burning at high temperatures.

\term{Opacity}: Mass extinction coefficient specifying the energy
fraction taken from a photon beam due to absorption and scattering
processes.

\term{Optical depth}: Measure of the level of transparency of a medium.
Optically thick (thin) refers to an opaque (transparent) medium.

\term{Radiative transfer}: Generic term used to describe the theory of
photon-matter interactions as radiation is transported through a
medium.

\term{Single degenerate (SD) scenario}: Progenitor channel for
\sneia\ in which a WD accretes material from a non-degenerate star.

\term{Supernova (SN)}: Energetic stellar explosion marking the end of
the life of some stars.

\term{Thermalization}: In \sneia, process by which $\gamma$-ray photons
deposit their energy in the ejecta and are reprocessed into thermal
radiation.

\term{Type Ia supernova (\snia)}: Supernova resulting from the
thermonuclear disruption of a carbon-oxygen white dwarf star.
  
\term{UVOIR luminosity}: Luminosity integrated over the full
ultraviolet-optical-infrared range (i.e., excluding escaping
high-energy radiation).
  
\term{White dwarf (WD) star}: Compact and dense star marking the
endpoint of the evolution of stars with a initial masses $M \lesssim
8$\,\msun.

\end{glossary}

\begin{BoxTypeA}[box]{Key points}
  \begin{itemize}
    
  \item Almost all progenitor scenarios for \sneia\ involve a
    carbon-oxygen white dwarf star in a binary system, but vary in the
    nature of the binary companion (non-degenerate star vs. another
    WD) and the explosion mechanism or combustion mode.
    \smallskip
    
  \item The energy input from the radioactive decay of
    \nifs\ synthesized in the explosion largely determines the
    luminosity evolution during the first 2--3 years.
    The spectroscopic evolution provides a time-dependent 
    scan of the chemically-stratified ejecta, and provides important
    clues related to its composition and expansion dynamics.
    \smallskip
    
  \item Most \sneia\ form a relatively homogeneous class, albeit with
    variations in their peak luminosity and spectroscopic
    properties. There are however events displaying peculiar light
    curves and spectra, and this observed diversity could indicate the
    existence of multiple progenitor channels.\smallskip
    
  \item While explosion models have different ejecta structures and
    levels of asymmetry, the predicted observational signatures are
    often similar and the modeling uncertainties still too large to
    make firm associations between individual \sneia\ and specific
    progenitor scenarios, except in a few rare cases. Long considered
    to be the most likely scenario, the explosion of
    near-Chandrasekhar-mass WDs still faces fundamental
    challenges and currently favored models involve pairs of sub-\mch\ WDs.
    \smallskip
    
  \item Future surveys will discover several hundreds of \sneia\ each
    night, mostly for cosmological studies. However, detailed 
    observations of individual events, from the pre-explosion phases
    to several years after, and covering the full electromagnetic
    spectrum (and soon gravitational waves), remain invaluable to
    constrain the mechanism(s) by which WD stars explode.
    
  \end{itemize}
\end{BoxTypeA}

\begin{glossary}[Nomenclature]
\begin{tabular}{@{}lp{34pc}@{}}
CE & common envelope\\
CSM & circumstellar material\\
DD & double degenerate\\
DDT & deflagration-to-detonation transition\\
DTD & delay-time distribution\\
EC & electron capture\\
HV & hypervelocity\\
HVF & high-velocity feature\\
IGE & iron-group elements\\
IME & intermediate-mass elements\\
LTE & local thermodynamic equilibrium\\  
MS & main sequence\\
NSE & nuclear statistical equilibrium\\
SD & single degenerate\\
\snia & Type Ia supernova\\
UVOIR & ultraviolet(UV)-optical-infrared(IR)\\
WD & white dwarf\\
\end{tabular}
\end{glossary}


\section{Introduction}\label{sect:intro}

Stars with initial masses $M \lesssim 8$\,\msun\ evolve to become
compact and dense objects known as white dwarfs (WD). Isolated WDs
slowly cool by radiating away their residual thermal energy on
timescales that can exceed the present age of the universe. However, a
carbon-oxygen (C-O) WD in a close binary system can accrete mass from
a non-WD star, or merge/collide with another WD, resulting in a
runaway thermonuclear fusion reaction chain that can unbind the entire
star.

Such stellar explosions, known as Type Ia supernovae (hereafter
\sneia), are among the most energetic in the universe, with typical
explosion energies of the order of $10^{44}$\,J. Their luminosity
reaches a typical peak value of $L_\mathrm{peak} \approx
10^{36}$\,\js\ (or $\sim 3 \times 10^9$\,\lsun), which is equivalent
to a sizable 20\%--30\% of the total luminosity of our own
galaxy. For this reason, \sneia\ are visible out to very large
distances, corresponding to a time when the universe was less than one
fifth of its present age, only $\sim2$\,Gyr after the Big Bang
\citep{Pierel/etal:2024}. Moreover, the realization that the light
curves of \sneia\ could be used to accurately calibrate their peak
brightness motivated the use of these events as distance indicators on
cosmological scales. This unique property allowed astronomers to use
\sneia\ as direct probes of the dynamical history of the universe, and
led to the discovery of its accelerated expansion.

Type Ia SNe are also of great astrophysical interest in their own
right.  As the main producers of iron in the universe ($\sim 2/3$ of
the iron content in our galaxy today; \citealt{Dwek:2016}),
\sneia\ are key players in the chemical evolution of galaxies.  They
inject a significant amount of kinetic energy and turbulence into the
interstellar medium, serving as a trigger for star formation and
driving galaxy-scale winds.  As an endpoint of binary stellar
evolution for low-mass stars, \sneia\ require an understanding of mass
transfer and accretion onto compact objects, and double-WD systems are
predicted to be the dominating source of background gravitational-wave
radiation for the future Laser Interferometer Space Antenna (LISA).
As is the case for other types of SNe, their long-lived remnants, some
of which are visible in our own galaxy (SN~1006, Tycho's SN~1572,
Kepler's SN~1604), are thought to be sites for cosmic-ray
acceleration.  Last, their explosion mechanism involves all the
complexity and richness of turbulent and shock-driven combustion, and
is therefore of interest beyond the astrophysical community.  Yet,
despite decades of observational and theoretical efforts, the exact
nature of the progenitors and explosion mechanisms of \sneia\ remains
an unsolved question to date.

In this chapter, we will first review the basic underlying physical
model for \sneia\ and characterize their observational properties. We
then give an overview of the main progenitor and explosion scenarios
invoked to explain the diversity of the \snia\ class. Last, we discuss
paths for future progress in the understanding of the
\snia\ phenomenon, both from an observational and theoretical
perspective.


\section{Basic physical picture and observational properties}\label{sect:basics}


\subsection{A carbon-oxygen white dwarf progenitor}

Unlike other types of supernovae resulting from the core collapse in
massive stars, there are no direct detections of the exploding star in
an \snia\ event, and only limited evidence related to the nature of
the binary companion (with firm non detections of a giant companion in
some cases). One therefore has to rely on indirect clues. One
such clue is the absence of lines due to hydrogen or helium --- the
two most abundant elements in the universe --- in their spectra around
maximum light.  This, combined with the occurrence of \sneia\ in
early-type galaxies dominated by old stellar populations, suggests
\sneia\ are associated with an evolved progenitor.  Last, the rapid
evolution of the light curve ($\lesssim 20$\,days from explosion to
peak luminosity, followed by a rapid decline on a similar timescale),
provides indirect evidence for a relatively compact progenitor.  Put
together, a carbon-oxygen WD star emerges as the only viable
progenitor for \sneia.

Carbon-oxygen white dwarfs are the endpoint of the evolution of stars
with initial masses $2 \lesssim M \lesssim 8$\,\msun\footnote{The
lower limit of $\sim2$\,\msun\ corresponds to the maximum initial mass
of a star that can lead to the formation of a helium WD in a close
binary system. For single stars this lower limit is slightly below
1\,\msun\ (i.e., our sun will eventually evolve to become a C-O
WD).}. Such stars reach core temperatures for the fusion of helium
into carbon and oxygen as they evolve off the main sequence (MS) to
become red giant stars. As the C-O core contracts, the gravitational
potential energy released is sufficient to eject the star's envelope,
and its density is such ($\sim 10^9$\,\kgcm\ on average) that the
electrons are degenerate and can be relativistic: It is this
degeneracy pressure of the electrons that counteracts the inward pull
of gravity, maintaining the star in hydrostatic equilibrium, before
the temperature threshold for carbon fusion is reached in the core.
Under these conditions, a WD star has a maximum mass it can sustain,
known as the Chandrasekhar mass \citep{Chandrasekhar:1931}, whose
value depends only on composition, and is approximately $\mch \approx
1.44$\,\msun\ for a cold, non-rotating, non-magnetic C-O WD star in
the ultra-relativistic limit:

\begin{equation}
  \mch = 0.197 \left( \frac{hc}{G} \right)^\frac{3}{2} \frac{1}{(\mu_\mathrm{e}
    m_\mathrm{H})} \approx 1.44\,\msun \mathrm{\quad for\quad } \mu_\mathrm{e}=2,
\end{equation}

\noindent
where $\mu_\mathrm{e}$ is the average molecular weight per electron
($\mu_\mathrm{e}=2$ for a fully ionized C-O mixture) and
$m_\mathrm{H}$ is the mass of the hydrogen atom (the other fundamental
constants have their usual meaning).

The existence of a maximum mass for WDs could explain why the majority
of \sneia\ appear to constitute a relatively homogeneous class, at
least by astronomical standards. Since the majority of C-O WDs are
born with an initial mass well below
1\,\msun\ \citep{Kepler/etal:2007}, this implies that they must
accrete mass from a binary companion, or merge with another WD in
order to reach
$\sim$1.4\,\msun\footnote{There are however models for
\sneia\ resulting from isolated WDs, such as the fission-based model
of \cite{Horowitz/Caplan:2021}.}.  However, the \snia\ class is
significantly more diverse than was originally thought, and
currently-favored models for the explosion involve WDs well below the
Chandrasekhar mass (sub-\mch).


\subsection{Basic explosion physics}\label{sect:expl}

In the classical model of \cite{Hoyle/Fowler:1960}, as the WD
approaches \mch, the core temperature (a few times $10^8$\,K) and
density ($\gtrsim 2\times 10^{12}$\,\kgcm) is sufficient to initiate
carbon fusion. Each subsequent increase in the WD mass via accretion
leads to a contraction of the core and an increase in temperature via
compressional heating, which in turn results in a rapid increase in
the nuclear reaction rate and a further rise in temperature. In a
normal (non-degenerate) star, this feedback process is self-regulated
via an increase in the gas thermal pressure which leads to an
expansion of the star and a drop of the core temperature. Under purely
degenerate conditions, however, the pressure is independent of
temperature. The above cycle is then self-sustained and the nuclear
fusion reactions may undergo a runaway process.

The increase in temperature is in part compensated via various cooling
mechanisms (neutrino emission, thermal conduction, and
convection). The WD will eventually expand when the thermal pressure
becomes comparable to the electron degeneracy pressure. Convective
motions can initially evacuate the surplus nuclear energy and regulate
nuclear reactions\footnote{This is known as the simmering phase and
can last for several thousands of years.}, but it is not efficient
enough to prevent the formation of a large temperature gradient and
the initiation of a deflagration via thermal conduction. The flame is
born, and a thermonuclear runaway follows.  Although this basic
physical picture differs somewhat in lower-mass (sub-\mch) WDs, the
explosion as a \snia\ always proceeds via a runaway thermonuclear C-O
fusion.

The fusion of $\sim 1.4$\,\msun\ of an equal mixture of $^{12}$C and
$^{16}$O into iron-group elements (IGEs) releases a nuclear
energy $E_\mathrm{nuc} \approx 2 \times 10^{44}$\,J,\footnote{In cgs
units this is equal to $2 \times 10^{51}$\,erg, where $10^{51}$\,erg
is sometimes referred to as one Bethe (in honor of Hans Bethe), or one
foe (fifty-one ergs).} which is sufficient to unbind the WD and
accelerate the ejecta to typical expansion velocities of the order of
$10^7$\,\ms, as inferred from the blueshifted absorption in spectral
lines:

\begin{equation}
  E_\mathrm{nuc} \approx -E_\mathrm{b} + E_\mathrm{kin},
\end{equation}

\noindent
where the gravitational binding energy (negative by convention)
$E_\mathrm{b} \approx -5 \times 10^{43}$\,J for a non-rotating
\mch\ WD, and the total kinetic energy $E_\mathrm{kin} \lesssim 1.5
\times 10^{44}$\,J.


\begin{figure}[t]
  \centering
  \includegraphics[width=0.75\textwidth]{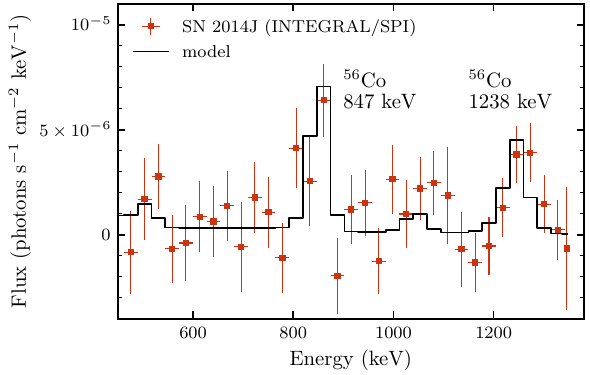} 
  \caption{Detection of the 847 and 1238\,keV ($1\,\mathrm{keV} \approx 1.6
    \times 10^{-16}$\,J) $\gamma$-ray decay
    lines of $^{56}$Co in INTEGRAL/SPI observations of the nearby Type
    Ia SN~2014J (data from
    \citealt{Churazov/etal:2014}). This is the only direct detection
    to date of $\gamma$-rays in a \snia\ event.}  
  \label{fig:co56}
\end{figure}

A few seconds after the onset of the explosion, the expansion follows
a self-similar regime referred to as homologous expansion.  The
expansion is such that adiabatic losses cause the WD to cool very
quickly: within 1 day after explosion, the WD has expanded from a
Earth-sized object ($R \approx 10^6$\,m) to an ejecta whose size is
similar to that of the solar system ($R \approx 10^{12}$\,m). This
would normally cause the maximum temperature in the ejecta to drop
from $\sim 10^9$\,K at a few seconds post explosion (at which point
all nuclear reactions have ceased) to a few 1000\,K at 1 day, and to
room temperature ($\sim 300$\,K) by 10\,days in the absence of an
additional source of energy.  \sneia\ would thus escape detection were
it not for the abundant production of a radioactive isotope of nickel,
\nifs\ \citep{Colgate/McKee:1969}. This isotope decays to \cofs\ via
electron capture (EC) and finally to \fefs\ via EC or positron decay
($\beta +$), emitting $\gamma$-rays and positrons whose energy is
deposited in the ejecta on timescales where adiabatic losses
(proportional to the differential change in volume, $\mathrm{d}V$)
become less important:

\begin{equation*}
{^{56}\mathrm{Ni}}\ \
\xrightarrow[t_{1/2} = 6.1\,\mathrm{days}]{Q_{\gamma} = 1.7\,\mathrm{MeV}}\ \ 
{^{56}\mathrm{Co} + \gamma + \nu_\mathrm{e}\ \mathrm{(EC)}}\ \ 
\xrightarrow[t_{1/2} = 77.2\,\mathrm{days}]{Q_{\gamma} = 3.6\,\mathrm{MeV}; Q_{\mathrm{e}^+} = 0.1\,\mathrm{MeV}} \ \ 
{^{56}\mathrm{Fe} + \gamma + \nu_\mathrm{e}\ \mathrm{(EC)}\quad
\mathrm{or}\quad ^{56}\mathrm{Fe} + \mathrm{e}^+ + \gamma + \nu_\mathrm{e}\ (\beta +)}
\end{equation*}

\noindent
where $Q_{\gamma}$ and $Q_{\mathrm{e}^+}$ correspond to the average energy per
decay ($1\,\mathrm{MeV} \approx 1.6 \times 10^{-13}$\,J) emitted in
$\gamma$-rays and positrons, respectively, and $t_{1/2}$ is the
isotope's half-life.  Only relatively recently has there been direct
observational evidence of this decay-powered luminosity in \sneia,
through the detection of $\gamma$-ray lines from \cofs\ in the nearest
\snia\ observed in the last 50 years, SN~2014J (Fig.~\ref{fig:co56}).

The products of nuclear burning largely depend on the local fuel
density. At sufficiently high densities, carbon fusion proceeds all
the way to the iron peak, and nuclear statistical equilibrium (NSE) is
established for temperatures $\gtrsim 5 \times 10^9$\,K. These
conditions ensure that the nucleosynthesis products do not depend on
individual reaction rates, but only depend on the peak temperature,
density, and electron fraction $\ye$. This latter quantity is set by
the initial metallicity of the WD (\ye\ is lower for higher
metallicity) and further modified by electron captures during the
high-density pre-explosion simmering phase and during the explosion
itself in \mch\ events. The final abundances in regions where NSE is
reached are determined in the subsequent freeze-out phase, in which free
particles reassemble into nuclei on timescales inversely proportional
to the density. In lower-density regions, carbon fusion does not
proceed to the iron peak as the silicon-burning process is
incomplete. These are the regions where most of the intermediate-mass
elements (IMEs) are synthesized.

The dependence of the nuclear burning products on density implies that
the mass of the exploding WD also affects the integrated yields, as
\mch\ WDs have larger central densities ($\rho_c \gtrsim
10^{12}$\,\kgcm) compared to lower-mass, sub-\mch\ WDs ($\rho_c
\lesssim 10^{11}$\,\kgcm). As a result, burning in NSE produces more
heavier neutron-rich stable isotopes (e.g., \feff\ and \nife) in
\mch\ explosions compared to sub-\mch\ explosions for a given initial
metallicity.  The stable nickel abundance can thus be used in
principle to distinguish \mch\ from sub-\mch\ explosions
\citep[e.g.,][]{Floers/etal:2020}.  This also applies to manganese,
whose solar abundance can only be explained if a significant fraction
of \sneia\ achieve the high-density burning conditions met in
\mch\ progenitors \citep{Seitenzahl/etal:2013b}.

\begin{figure}[t]
  \centering
  \subfloat[$t=1$\,s]{\includegraphics[width=.425\textwidth]{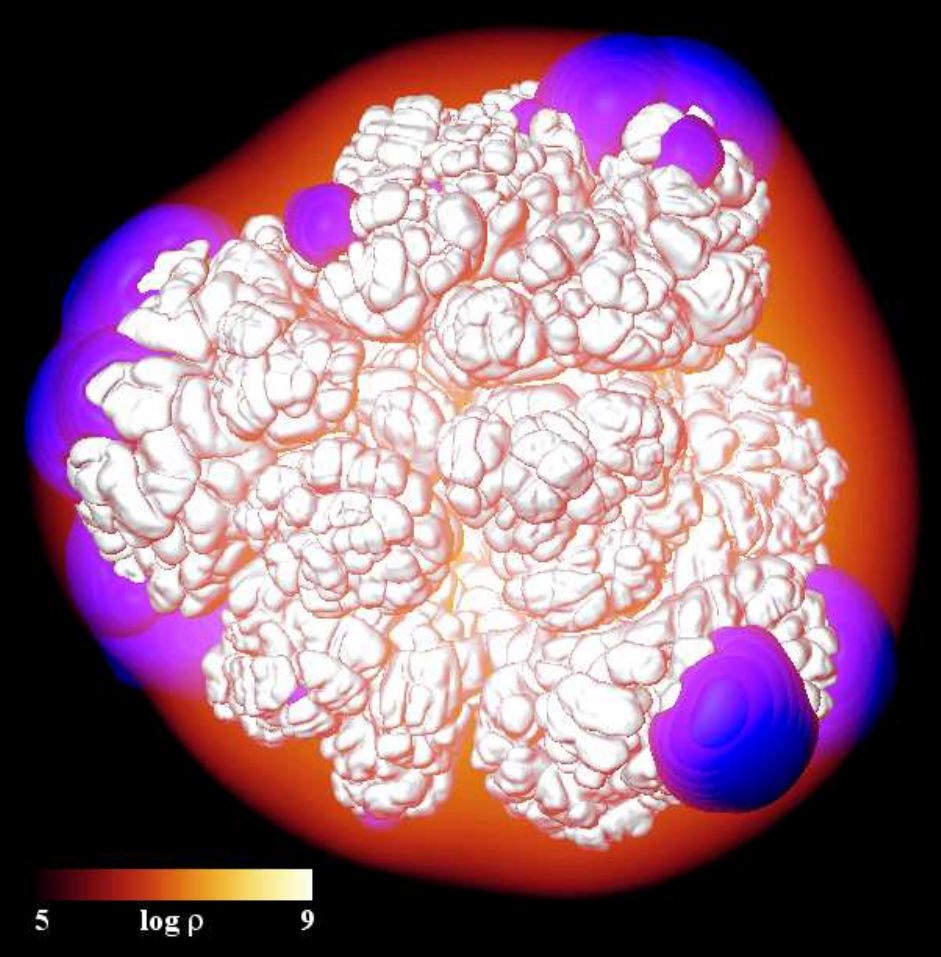}}\hspace{.02\textwidth}
  \subfloat[$t=100$\,s]{\includegraphics[width=.55\textwidth]{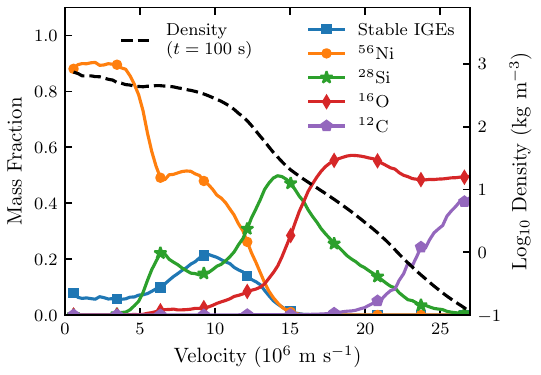}}\\
  \vspace{-0.25cm}
  \subfloat[Stable IGEs]{\includegraphics[width=.2\textwidth]{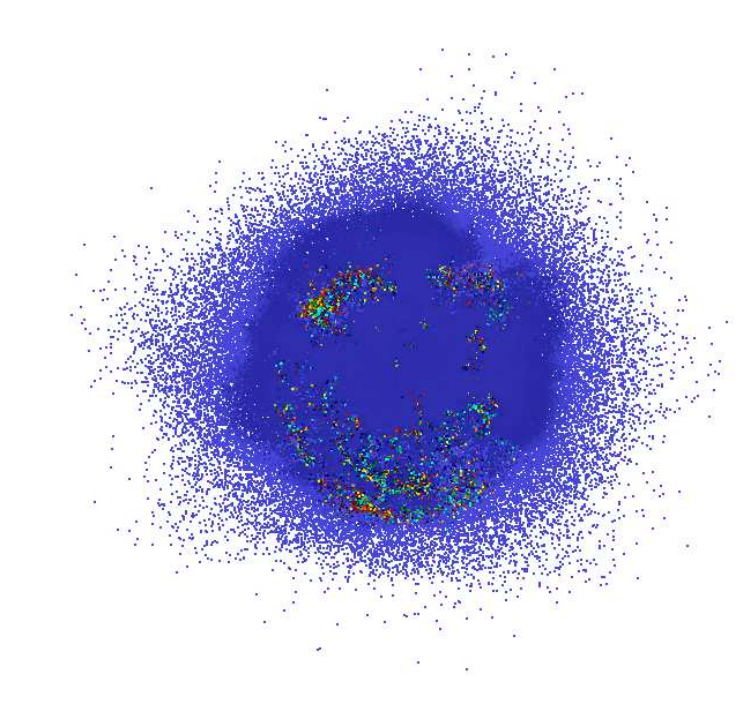}}
  \subfloat[\nifs]{\includegraphics[width=.2\textwidth]{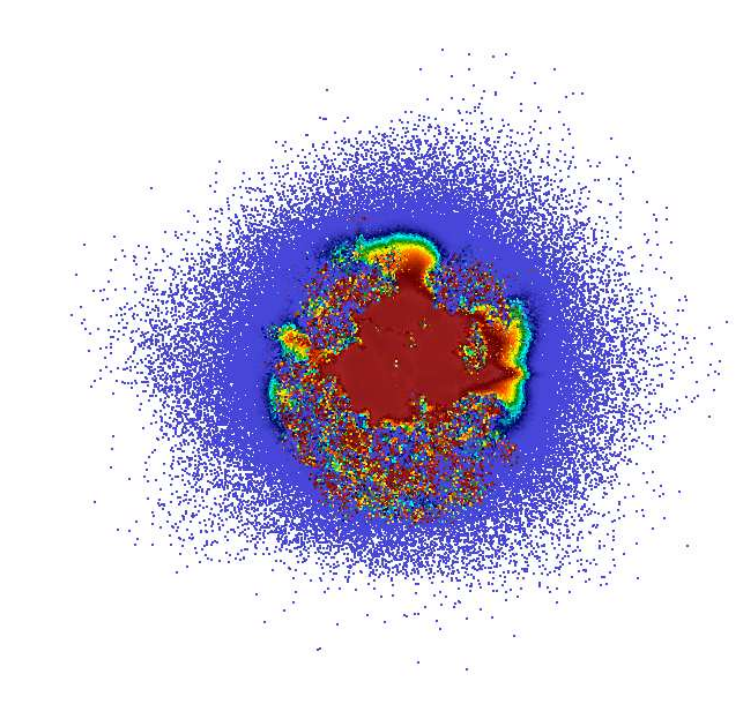}}
  \subfloat[$^{28}$Si]{\includegraphics[width=.2\textwidth]{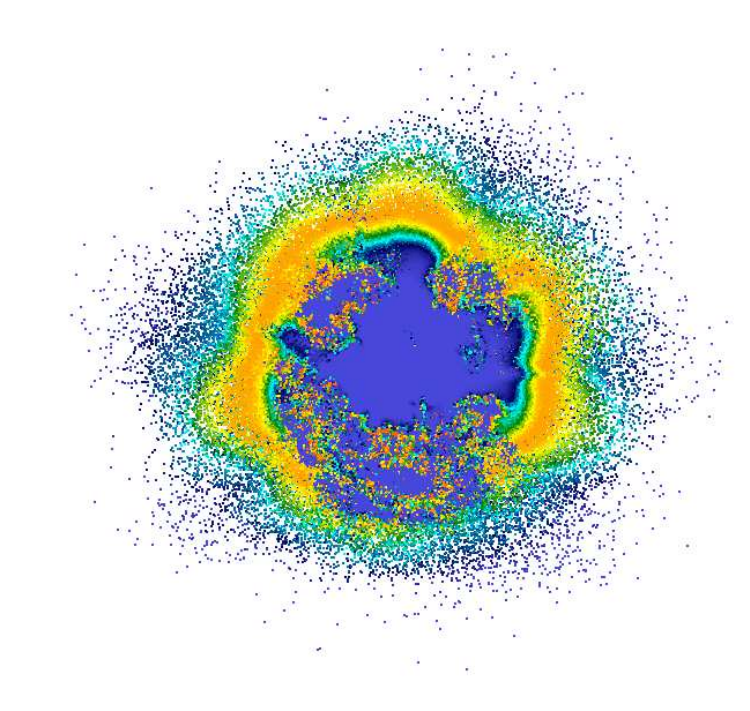}}
  \subfloat[$^{16}$O]{\includegraphics[width=.2\textwidth]{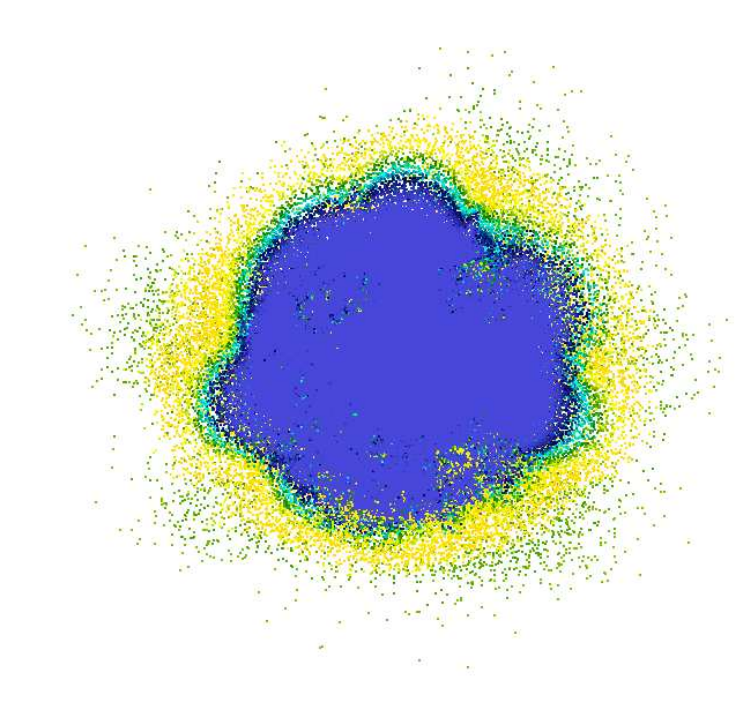}}
  \subfloat[$^{12}$C]{\includegraphics[width=.2\textwidth]{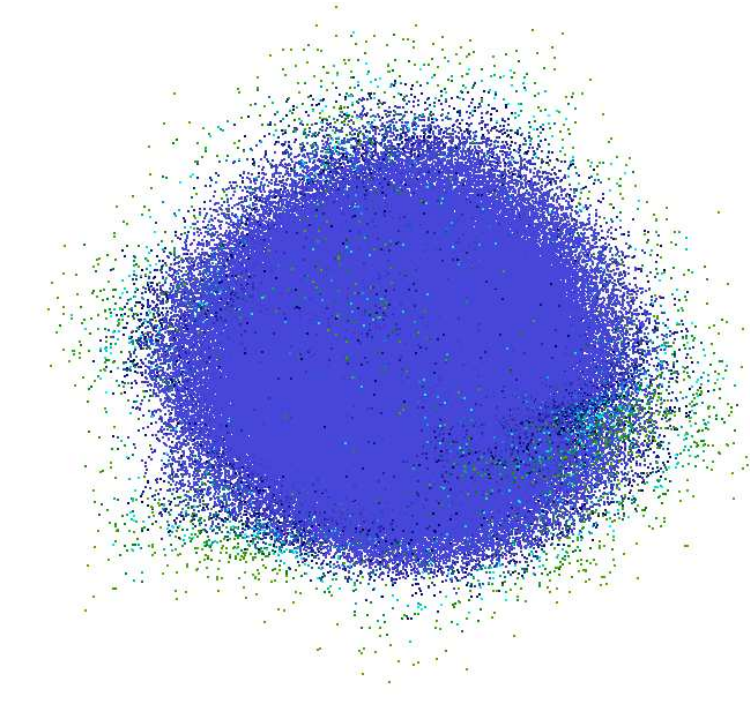}}\\
  \vspace{0.2cm}
  {\includegraphics[width=0.35\textwidth]{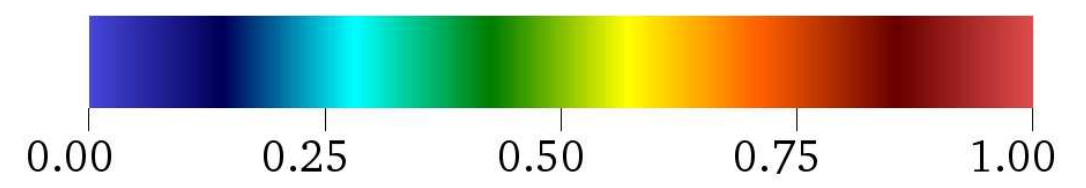}\\Mass Fraction}
  \vspace{0.25cm}
  \caption{Example of a 3D \mch\ delayed-detonation model (N100;
    \citealt{Seitenzahl/etal:2013}).
    (a) Density (in $\gcc = 10^3$\,\kgcm; orange color scale),
    deflagration plumes (white), and detonation front (blue) at
    $t=1$\,s post explosion.
    (b) Spherically averaged density (dashed line) and composition
    profiles for representative isotopes (solid lines) at $t=100$\,s
    post explosion. The 2D-projected distribution of each isotope is
    shown in the bottom row in panels (c)--(g). The graphics shown in
    panels (a) and (c)--(g) are reproduced with permission from
    \citet[their Figs.~2 and 4]{Seitenzahl/etal:2013}.
  }
\label{fig:n100}
\end{figure}

Together with the chemical stratification inferred from the time
evolution of \snia\ spectra (see next section), we can already
highlight three basic ingredients for a successful \snia\ model (see
box below). The final ingredient of course is to allow for some
diversity to match what is seen in nature. One possible model that
satisfies these requirements is illustrated in Fig.~\ref{fig:n100}.

\begin{BoxTypeA}[box:minreq]{Three basic ingredients for a successful \snia\ model}
\begin{enumerate}
    
\item \textbf{Explosion energy:} of the order of $10^{44}$\,J to
  unbind the WD and accelerate the ejecta to characteristic velocities
  of the order of $10^7$\,\ms. This implies the WD must be (almost)
  entirely burnt to release sufficient nuclear energy from fusion of
  C-O into heavier elements.\\
  \textbf{Note:} some explosion models (e.g., pure deflagrations)
  result in too low an explosion energy to completely unbind the WD,
  in which case a bound remnant is left behind. Such models have been
  found to be a good candidate for low-luminosity Type Iax events.
  \smallskip
    
\item \textbf{Nucleosynthesis:} a \nifs\ yield in the range
  0.1--1.0\,\msun\ (to power the light curve) + several
  0.1\,\msun\ of IMEs (to match the spectra) + some stable IGE
  isotopes (\nife, \feff) + possible traces of unburnt C-O. This
  requires thermonuclear burning at a large range of densities.
  \smallskip
  
\item \textbf{Chemical stratification:} stable IGEs + \nifs\ near the
  center and IMEs in the outer layers. This requires a large
  (exponentially decreasing) density gradient and weak large-scale
  mixing during the explosive phase.\\
  \textbf{Note:} the inner ejecta of violent merger models are
  dominated by the ashes of the incompletely burned secondary WD (O,
  Ne, Mg), and pure deflagration models predict significant amounts of
  unburnt C-O material as a result of large-scale mixing during the
  explosion.
    
\end{enumerate}
\end{BoxTypeA}

\subsection{Luminosity and spectroscopic evolution}

\begin{figure}[t]
  \centering
  \includegraphics[width=\textwidth]{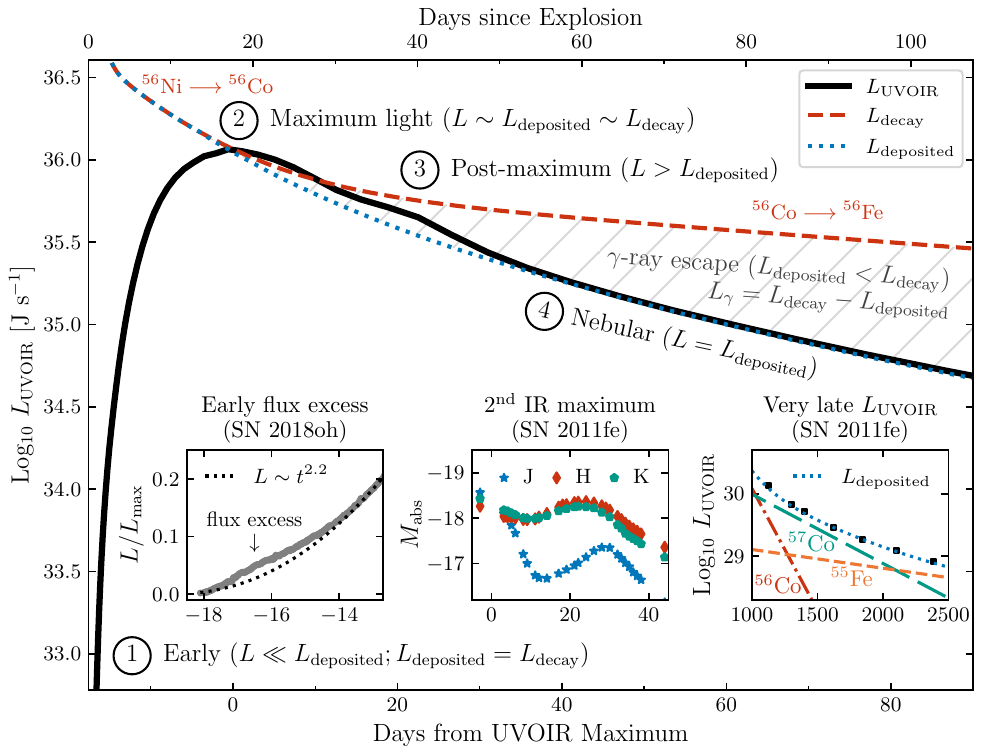}
  \caption{Luminosity evolution for a \snia\ model with $\sim
    0.5$\,\msun\ of \nifs\ (black solid line; DDC15 model of
    \citealt{Blondin/etal:2015}). The dashed and dotted lines show the
    decay luminosity and the deposited decay power, respectively. The
    hatched area highlights the increasing $\gamma$-ray escape past
    maximum light. The insets give examples of early flux excess
    (left; data from \citealt{Dimitriadis/etal:2019}), the secondary near-IR
    maximum (middle; data from \citealt{Matheson/etal:2012}), and
    the contribution of additional decay chains to the luminosity at
    very late times (right; data from \citealt{Tucker/etal:2022}). The
    axis in each case corresponds to days from maximum
    light.}
  \label{fig:lbol}
\end{figure}

We illustrate the evolution of the UVOIR luminosity for a \snia\ model
in Fig.~\ref{fig:lbol}. As the ejecta expands, the radiation emerges
from deeper ejecta layers\footnote{At least in a Lagrangian sense, as
these deeper layers are expanding at several $10^6$\,\ms.}, offering a
time-dependent scan of its chemical structure
(Fig.~\ref{fig:specevol}).  In what follows we will discuss four main
phases of the evolution over the first few months post explosion,
focusing on normal \snia\ events.

\subsubsection{Early evolution}\label{sect:early}

Although \nifs\ decay is the main driver for the luminosity evolution,
the early luminosity of \sneia\ can be affected by other factors.
Most \snia\ explosion models involve combustion driven by a strong
supersonic shock (i.e., a detonation). A predicted observational
consequence is a UV-optical flash of radiation resulting from the
shock-heated stellar envelope and lasting $\lesssim 1$\,h, whose
luminosity $L_\mathrm{shock} \approx 10^{32}$\,\js\ (i.e., four orders
of magnitude lower than the peak luminosity;
\citealt{Rabinak/etal:2012}) has never been observed\footnote{However,
it is the non-detection of this signature in the nearby Type Ia
SN~2011fe that enabled to place a strong constraint on the progenitor
radius $R_\star \lesssim 0.02$\,\rsun\ \citep{Bloom/etal:2012}. Also
note that an even earlier, shorter-lived signature is expected in
X-rays, when the shock breaks out from the stellar envelope.}. This
early-time flash can in principle be followed by a dark phase during
which \nifs\ decay photons are still trapped within the ejecta
(estimated to last $\sim 1$\,day in the case of SN~2011fe;
\citealt{Mazzali/etal:2014}).

During the first few days that follow, the $\gamma$-ray photons
emitted in the \nifs\ decay chain are completely thermalized by
Compton scattering off free electrons. Due to the high initial optical
depth, this thermal radiation diffuses over a time scale significantly
larger than the time elapsed since the explosion, such that the
initial luminosity is very low. As the ejecta expands, the photon
escape time becomes less than the characteristic expansion time, and
the luminosity increases. This initial increase is often modeled as a
power law in time, $L(t) \propto t^\alpha$, where $\alpha=2$ matches well
the early light curve of SN~2011fe \citep{Nugent/etal:2011}. However,
both theoretical predictions and observations find significant
deviations from $\alpha=2$ \citep{Piro/Nakar:2014,Firth/etal:2015}.
Moreover, an early flux excess has been observed in several
\sneia\ (left inset in Fig.~\ref{fig:lbol}), which could result from
the interaction with a binary companion or circumstellar material
(CSM), or from the presence of \nifs\ in the outer regions of the
ejecta.

\begin{figure}[t]
  \centering
  \includegraphics[width=0.6\textwidth]{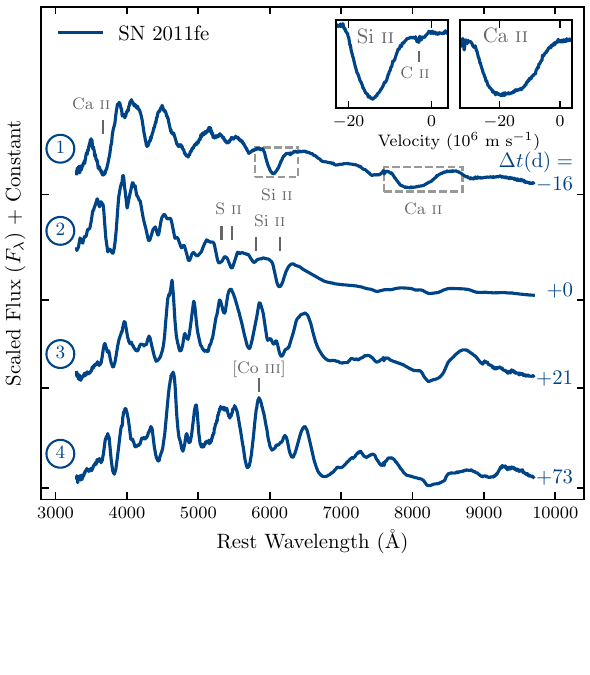}
  \vspace{-1.9cm}
  \caption{Spectroscopic evolution of the Type Ia SN~2011fe over the
    first $\sim$100 days post explosion (data from
    \citealt{Pereira/etal:2013}). The right labels give the 
    spectral phase in days from maximum light, and the circled numbers
    on the left correspond to the phases highlighted in
    Fig.~\ref{fig:lbol}. The tickmarks on the Y-axis give the
    zero-flux level of each spectrum. A few key spectroscopic
    features are highlighted in gray. The insets show examples of
    high-velocity features at early times in the \ion{Si}{2}~6355\,\AA\ and
    \ion{Ca}{2} near-IR triplet line profiles. A small absorption due to
    \ion{C}{2}~6580\,\AA\ (i.e., unburnt carbon) can also be seen in
    the left inset.}
  \label{fig:specevol}
\end{figure}

The optical spectra at these times are dominated by Doppler-broadened
lines of ions from IMEs synthesized in the explosion
(e.g., \ion{Si}{2}, \ion{S}{2}, \ion{Ca}{2} etc.), as well as lines
from unburnt carbon (first spectrum in Fig.~\ref{fig:specevol}). The
characteristic P-Cygni profile shape of these lines displays a
prominent blueshifted absorption that is used to infer the expansion
velocities of the corresponding ion. In particular, high-velocity
features (HVFs) are routinely found in lines of \ion{Ca}{2} and
\ion{Si}{2} (see insets in Fig.~\ref{fig:specevol}) These HVFs are
sometimes associated with a spectropolarimetric signature, indicating
an asymmetric distribution of the emitting material
\citep[e.g.][]{Kasen/etal:2003}.


\subsubsection{Peak luminosity (maximum light)}

The initial increase in luminosity is counterbalanced by the
exponentially decreasing decay power, which causes a maximum in the
light curve. At this time, the UVOIR luminosity is comparable (but not
physically equivalent) to the decay power injected in the inner ejecta
layers: this is known as Arnett's rule \citep{Arnett:1979}, and can be
formally expressed as $L_\mathrm{UVOIR}(t_\mathrm{peak}) \approx
\alpha L_\mathrm{decay}(t_\mathrm{peak})$, with $\alpha \approx 1$.
Several studies have found this rule to be accurate to within
10\%--20\,\%\ \citep[e.g.][]{Blondin/etal:2013}.  The
maximum-light spectra are still dominated by lines from IMEs, and
early-time signatures of unburnt carbon have usually disappeared by
then.


\subsubsection{Post-maximum evolution}

Two notable phenomena occur after the peak in luminosity. First, the
rate of energy actually deposited in the ejecta (dotted curve in
Fig.~\ref{fig:lbol}) becomes an increasingly smaller fraction of the
radioactive decay power (mainly from \cofs\ decay at this time; dashed
curve). This is due to the increase in the mean free path of
$\gamma$-ray photons, an ever-increasing fraction of which manages to
escape directly from the ejecta without interaction\footnote{The
characteristic $\gamma$-ray escape time scales with the square root of
the total ejecta mass \citep{Jeffery:1999}, and has led several
authors to conclude that a significant fraction of \sneia\ result from
sub-\mch\ progenitors
\citep{Stritzinger/etal:2006,Scalzo/etal:2014,Wygoda/etal:2019}.}. Second,
the luminosity exceeds the energy deposition rate during $\sim
30$\,days, which corresponds to the time needed to radiate the energy
stored at previous times. Around 40 days post explosion, we even
observe a slight inflection in the light curve: radiative processes
lead to a cooling of the ejecta, including emission by forbidden lines
(such as [\ion{Co}{3}]\,5888\,\AA; \citealt{Dessart/etal:2014c}). This
releases a significant amount of radiative energy ($\Delta
E_\mathrm{rad} \approx a_R \Delta T^4 \Delta V$, where $a_R$ is the
radiation constant), of the order of $10^{35}$\,\js. To this inflection
in the evolution of the UVOIR luminosity corresponds a second peak in
the light curves in the near-IR bands ($JHK$; see middle inset in
Fig.~\ref{fig:lbol}).

More luminous \sneia\ decline more slowly past maximum light in
several photometric bands (i.e., their light curves are broader)
compared to lower-luminosity events: this is known as the
width-luminosity relation \citep{Phillips:1993}. It results mostly from the
faster evolution to redder colors in lower-luminosity events around
maximum light, which causes a more rapid decline in the blue optical
bands \citep{Kasen/Woosley:2007}.


\subsubsection{Late-time evolution: the nebular phase}\label{sect:neb}

Finally, from $\sim 50$\,days after the explosion onward, the ejecta
becomes almost completely transparent, and the luminosity equals (and
is physically equivalent to) the rate of decay energy deposition in the
ejecta, an increasing fraction of which comes from positrons from
\cofs\ decay\footnote{Other radioactive decay chains become important
at even later times, such as $^{57}$Co ($t_{1/2}\approx 272$\,days) and
$^{55}$Fe ($t_{1/2}\approx 1000$\,days), both confirmed
observationally in the nearby SN~2011fe (see right inset in
Fig.~\ref{fig:lbol}).}: this is known as the nebular phase.  One can
further show that the time-weighted integral of the UVOIR luminosity
is equal to that of the deposited decay power by this time
\citep{Katz:2013}:

\begin{equation}
  \label{eqn:qtildekatzuvoir}
  \int_0^t t' L_\mathrm{UVOIR}(t')\ \mathrm{d}t' = 
  \int_0^t t' L_\mathrm{dep}(t')\ \mathrm{d}t' \qquad \mathrm{for}\ t
  \gg t_\mathrm{peak},
\end{equation}

\noindent
a property that can be used to infer the \nifs\ mass from observations
without relying on the less accurate Arnett rule.

\begin{figure}[t]
  \centering
  \includegraphics[width=\textwidth]{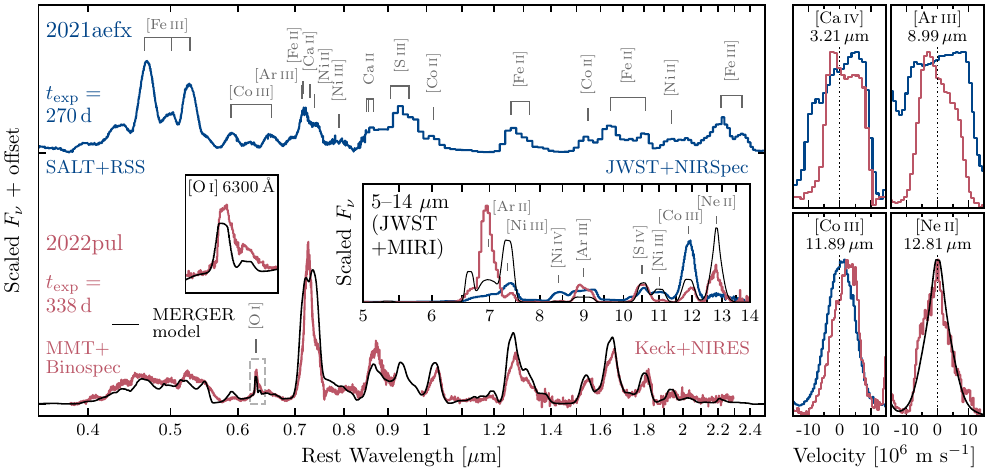}\hspace*{0.0cm}
  \caption{\textbf{Left panel:} Nebular spectra of SN~2021aefx at 270\,days post
    explosion (top) and of SN~2022pul at 338\,days post explosion
    (bottom). The SN~2022pul spectrum is compared to a violent merger model,
    which correctly predicts a narrow
    [\ion{O}{1}]\,6300\,\AA\ doublet (solid line; see also left inset).
    \textbf{Right panels:}
    Comparison of mid-IR line profiles in both spectra, normalized to
    their peak intensity. Lines
    in this range are more isolated and
    reveal variations in morphology (boxy vs. peaked), as well as
    kinematic offsets of several $10^6$\,\ms. The mid-IR spectrum of
    SN~2022pul shows a prominent line due to
    [\ion{Ne}{2}]\,12.81\,\micron, which is characteristic of the
    violent merger of two WDs \citep{Kwok/etal:2024}.}
  \label{fig:nebspec}
\end{figure}

The nebular spectra now probe the innermost ejecta and reflect the
iron-group dominated composition there (Fig.~\ref{fig:nebspec}). The
most prominent lines are due to forbidden transitions of Fe and Co,
with some evidence of weak lines from nickel. The overlying IMEs,
visible at earlier phases, also contribute to the cooling.  Since all
of the \nifs\ synthesized in the explosion has long decayed by this
time, all the nickel left in the ejecta is made up of stable isotopes
(mainly \nife).  As noted previously, the abundance of
these stable isotopes can in principle be used to constrain the mass
of the exploding WD.

The forbidden lines in nebular \snia\ spectra are optically thin and
hence only visible in emission. The morphology of their profiles can
be used to constrain the chemical distribution and ejecta geometry
(right panels in Fig.~\ref{fig:nebspec}). For instance, several
\sneia\ display velocity shifts in their nebular line profiles
(i.e., the lines are not centered on their rest wavelength), that could
result from an off-center explosion \citep{Maeda/etal:2010a}. Other
events display a horn-like double-peaked morphology, consistent with
predictions of WD-WD collision models \citep{Dong/etal:2015}. More
recently, the James Webb Space Telescope (JWST) has paved the way for
systematic studies of \sneia\ in the mid-IR spectral
range. Observations of one event, SN~2022pul, have revealed the
presence of a strong line due to [\ion{Ne}{2}]\,12.8\,\micron,
presented as a smoking-gun signature of the violent merger of two WDs 
(see MERGER model spectrum in Fig.~\ref{fig:nebspec}).

\subsection{Observational diversity}

\begin{figure}[t]
  \centering
  \includegraphics[width=.4\textwidth]{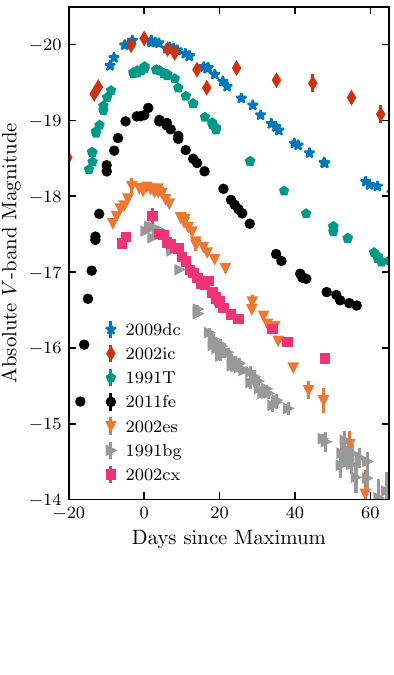}\hspace*{0.0cm}
  \includegraphics[width=.6\textwidth]{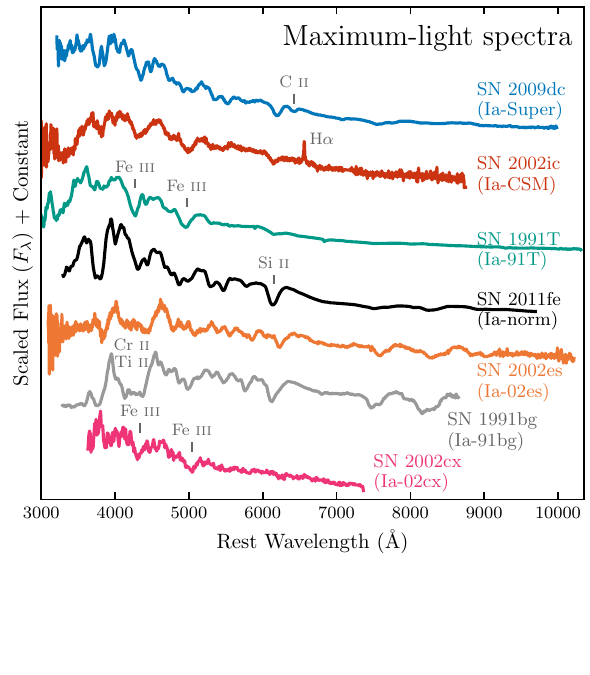}
  \vspace{-2.2cm}
  \caption{The observational diversity of \sneia, both in terms of
    their peak luminosity and light-curve shape (here in the $V$ band;
    left panel) and their spectroscopic properties (here around maximum
    light; right panel). Data sources:
    SN~2009dc \citep{Taubenberger/etal:2011},
    SN~2002ic \citep{Hamuy/etal:2003,Wood-Vasey/etal:2004},
    SN~1991T \citep{Phillips/etal:1992,Lira/etal:1998,Altavilla/etal:2004},
    SN~2011fe \citep{Richmond/Smith:2012,Pereira/etal:2013},
    SN~2002es \citep{Hicken/etal:2009a,Ganeshalingam/etal:2012},
    SN~1991bg \citep{Filippenko/etal:1992b,Leibundgut/etal:1993,Turatto/etal:1996},
    SN~2002cx \citep{Li/etal:2003}.
  }
  \label{fig:pec}
\end{figure}

The basic physical picture outlined above applies to the bulk of the
\snia\ population, and can accommodate part of the observed diversity
(in rise time, peak luminosity, post-maximum decline rate, color
evolution, and spectroscopic properties; see Fig.~\ref{fig:pec}), some
of which could be simply due to viewing-angle effects in an asymmetric
explosion.  At the high-luminosity end are the 91T-like \sneia\ that
resemble the prototypical SN~1991T, characterized by a higher
ionization and abundance of IGEs in the outer layers, and sometimes
associated with CSM interaction (see \citealt{Phillips/etal:2024} for
a recent review). At the low-luminosity end, 91bg-like
\sneia\ (resembling the prototypical SN~1991bg) are characterized by a
deep \ion{Ti}{2}/\ion{Cr}{2} absorption trough in their early-time
optical spectra, redder colors at early times, and the absence of a
secondary maximum in their near-IR light curves. Some authors have
explained this diversity as a continuous sequence in
temperature/ionization related to the \nifs\ yield
\citep[e.g.][]{Nugent/etal:1995}.

There are however peculiar events that possibly constitute distinct
sub-classes, and these are sometimes collectively referred to as
``thermonuclear supernovae'' or ``white dwarf supernovae'' (see
\citealt{Taubenberger:2017} for a review). Of these, we highlight
super-luminous \sneia\ (e.g., SN~2009dc) sometimes associated with the
explosion of rapidly-rotating super-massive WDs
\citep{Yoon/Langer:2005}\footnote{Often referred to as ``super-\mch''
\sneia\ since their mass exceeds $\sim 1.4$\,\msun, although this
canonical value only applies to a non-rotating WD, which is not the
case here.}, which could also form via the merger of a sub-\mch\ WD
and the C-O core of an asymptotic giant branch star inside a common
envelope (core-degenerate scenario; \citealt{Kashi/Soker:2011});
\sneia\ that display strong narrow hydrogen lines resulting from
interaction with a dense CSM (Ia-CSM; e.g., SN~2002ic); low-luminosity
events with low ejecta velocities that resemble SN~2002cx (referred to
as Type Iax SNe; see \citealt{Jha:2017} for a review); and
low-luminosity events similar to SN~2002es with a slow post-maximum
decline in luminosity (i.e., contrary to the width-luminosity
relation). Example light curves and maximum-light spectra of these
events are shown in Fig.~\ref{fig:pec}.

\subsection{Modeling challenges}

Before presenting possible progenitor scenarios and explosion
mechanisms in the following section, we first comment on some of the
theoretical and numerical challenges associated with modeling \sneia.

\subsubsection{Thermonuclear burning fronts}

The first of these concerns modeling thermonuclear burning fronts (see
\citealt{Roepke:2017} for a review). Combustion fronts can propagate
via two modes: subsonic deflagrations (which propagate via thermal
conduction and are accelerated via interaction with turbulence) and
supersonic detonations (which propagate via a strong shock).  A
deflagration can also transition to a detonation via a
deflagration-to-detonation transition (DDT), relevant for
\mch\ \snia\ models, but which poses a
major challenge even in terrestrial combustion experiments
\citep{Poludnenko/etal:2019}.

Because of the steep dependence of nuclear reaction rates on
temperature, combustion fronts are confined to a thin layer that can
be well below $10^{-3}$\,m, which is more than nine orders of
magnitude smaller than the WD radius ($\gtrsim 10^6$\,m). The modeling
of deflagrations requires knowledge of microphysical processes that
cannot be resolved in multi-dimensional simulations, even with
advanced numerical methods such as adaptive mesh refinement. While
such knowledge is not necessary when modeling detonation fronts, these
display a multi-dimensional cellular pattern whose inner structure is
also left unresolved. In both cases, the combustion front must either
be artificially broadened to match the resolution of the numerical
grid, or treated as a sharp discontinuity (level-set approach; as seen
in Fig.~\ref{fig:n100}) whose propagation speed is determined a priori
as opposed to computed self consistently.

\subsubsection{Explosive nucleosynthesis}

Computing accurate yields in \snia\ explosion simulations is also
faced with a resolution problem, as the reaction length scales vary by
several orders of magnitude (see \citealt{Seitenzahl/Townsley:2017}
for a review). The number of isotopes involved in the nuclear
reactions relevant to carbon burning is also prohibitively large to be
included in multi-dimensional hydrodynamical simulations of the
explosion (200--300 isotopes in typical \snia\ reaction networks). The
strategy is then to consider smaller networks coupled to the
hydrodynamics that match the nuclear energy release --- which serves
for instance to propagate the detonation front --- predicted using
larger networks, and then compute the accurate nucleosynthesis as part
of a post-processing step (using Lagrangian tracer particles;
e.g., \citealt{Travaglio/etal:2004}).

\subsubsection{Radiative transfer}

To predict the observable signatures from a given explosion model, one
needs to model the interaction of the radiation with the
rapidly-expanding ejecta, i.e., solve a radiative-transfer (RT)
problem. There are various approaches to this, from solving the RT
equation along rays to following photon packets in a Monte Carlo
experiment.  Most RT simulations are initiated at $\sim 1$\,day after
the explosion, at which point the ejecta are in homologous expansion
and one can choose to ignore the hydrodynamical part of the problem.
Up until the nebular phase, however, time-dependent effects should be
taken into account since there is a delay between the injection of
energy from the \nifs\ decay chain and the resulting escape of
thermalized photons from the SN ejecta. Non-local energy deposition by
$\gamma$-rays and non-thermal effects resulting from fast electrons
generated in Compton scatterings of these $\gamma$-rays also need to
be treated self consistently.

Furthermore, RT simulations in the context of \sneia\ are complicated
due to the nature of the opacity
\citep[e.g.][]{Pinto/Eastman:2000b}. The dominance of atomic lines
(bound-bound transitions), which both absorb and scatter incoming
photons, has the effect of decoupling the radiation field from the
local thermodynamic conditions of the gas (temperature and electron
density). One then needs to solve numerous rate equations to compute
the populations $n_i$ (number density, m$^{-3}$) of each atomic level
$i$ for each ion considered in the solution. The IGEs that dominate
the composition of \snia\ ejecta have several thousands of individual
levels and millions of bound-bound transitions between them. Solving
the complete set of rate equations comes at a huge computational cost,
even in 1D, which becomes prohibitively large in 3D.

One strategy is to treat the entire problem assuming spherical
symmetry (1D), but this overlooks the inversion of the chemical
stratification between the stable IGEs and \nifs-rich layers as
predicted in 3D models of the explosion (for \mch\ DDT
models). Another approach is to start with spherical averages of a 3D
explosion model, but this then artificially mixes together regions
burnt at a wide range of densities and hence leads to chemical
inconsistencies \citep[e.g.][]{Pakmor/etal:2024}. Last, treating the
entire problem (hydrodynamics + RT) in 3D requires the sacrifice of
physical consistency (e.g., in the treatment of ionization) to save
computational costs. Even in 1D, starting with the same initial ejecta
model does not guarantee that different codes will yield the same
results \citep{Blondin/etal:2022b}.


\section{Progenitor scenarios and explosion mechanisms}\label{sect:prog}

\begin{figure}[t]
  \centering
  \includegraphics[width=\textwidth]{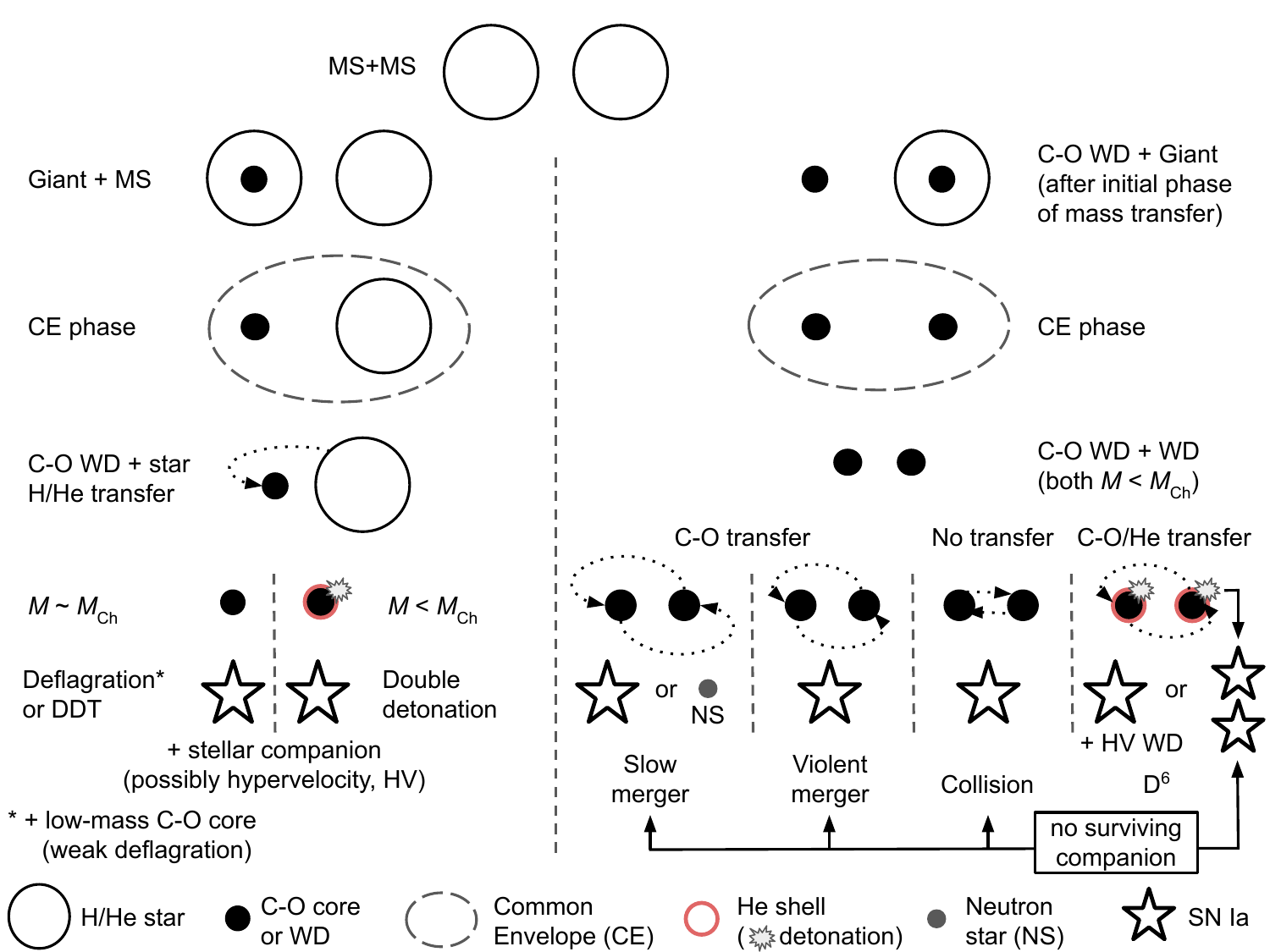}
  \vspace{0cm}
  \caption{Schematic view of the six \snia\ progenitor
    channels presented in this chapter, starting with a main sequence
    (MS) binary system and proceeding through different evolutionary
    stages and final outcomes (each separated by a dashed vertical
    line).}
  \label{fig:prog}
\end{figure}

There are several viable progenitor scenarios for \sneia, with
specific explosion mechanisms associated with each. For a long time,
the explosion of a near-\mch\ WD was considered the most likely
candidate, at least from a theoretical standpoint, but recent years
have witnessed a regained interest in models involving sub-\mch\ WD
pairs, in particular violent mergers and double-detonation
models. Different models, however, can yield similar observable
signatures, as the outcome of the explosion in terms of ejecta
dynamics and composition are sometimes weakly sensitive to the initial
state of the WD or the details of the nuclear
burning\footnote{P.~Hoeflich coined the term ``stellar amnesia'' to
describe this lack of sensitivity to the initial conditions
\cite[e.g.][]{Hoeflich:2006}.}. That said, a few observed \sneia\ have
displayed smoking-gun signatures associated with a particular
scenario. In what follows we review the main explosion models
considered by the community today. The different progenitor channels
associated with each model are schematically illustrated in
Fig.~\ref{fig:prog}, and Table~\ref{tab:prog} summarizes some of their
most salient features.


\begin{table}[t]
\centering
\caption{Basic characteristics of progenitor scenarios for \sneia.}\label{tab:prog}
\begin{threeparttable}
\begin{tabular}{l@{\hspace{1.8mm}}cccccc}
\toprule
\multicolumn{1}{c}{\textit{Scenario}}            & \multicolumn{1}{c}{\textit{Combustion}}      & \multicolumn{1}{c}{\textit{$M_\mathrm{WD}$}} & \multicolumn{1}{c}{\textit{Companion}} & \multicolumn{1}{c}{\textit{Surviving}}            & \multicolumn{1}{c}{\textit{Early}}        & \multicolumn{1}{c}{\textit{Nebular}}          \\
                                        & \multicolumn{1}{c}{\textit{mode}\tnote{\,a}} & \multicolumn{1}{c}{\textit{(primary)}}                       &                                  & \multicolumn{1}{c}{\textit{companion}\tnote{\,b}} & \multicolumn{1}{c}{\textit{interaction}}  & \multicolumn{1}{c}{\textit{lines}\tnote{\,c}} \\
\colrule
\multicolumn{7}{c}{\it Single WD progenitor systems} \\
\colrule
\multirow{2}{*}{\makecell{\\[0.3em]Single degenerate}} & Delayed-DET     & \multirow{2}{*}{\makecell{\\[0.3em]$\sim\mch$}}         & \multirow{2}{*}{\makecell{\\[-0.5em]MS/Giant\\or He star}}      & \multirow{2}{*}{\makecell{HV star\\or slow stripped Giant\tnote{\,d}\\(+ low-mass C/O core\tnote{\,e}\,\,\,)}}                       & \multirow{2}{*}{{\makecell{\\[-0.5em]Companion\\and CSM}}}    & \makecell{narrow H$\alpha$\\+ velocity shifts\tnote{\,f}} \\ \cmidrule(r){2-2}\cmidrule(l){7-7}
                                        & DEF                            &                                     &                                &           &                               & narrow [\ion{O}{1}]                \\
\colrule
Double detonation                       & DET                            & Sub-\mch                            & He star                        & HV star                       & \multirow{2}{*}{\makecell{Companion\\and CSM}}   & $\cdots$                           \\
                                        & (He+C)                         &                                     &                                &                               &                                       &                                    \\
\colrule
\multicolumn{7}{c}{\it Double WD progenitor systems} \\
\colrule
Slow WD merger                          & DET                            & Sub-\mch                            & Sub-\mch\ WD                   & None                          & C-O envelope                          & $\cdots$                           \\
(Double-degenerate)                     &                                & or $\sim\mch$                       &                                & or NS\tnote{\,g}                         &                                       &                                    \\
\colrule
Violent WD merger                       & DET                            & Sub-\mch                            & Sub-\mch\ WD                   & None                          & $\cdots$                              & [\ion{Ne}{2}]                      \\
                                        &                                &                                     &                                &                               &                                       & + narrow [\ion{O}{1}]                \\
\colrule
D$^6$ (one explosion)                   & \multirow{2}{*}{\makecell{\\[-0.8em]DET\\(He+C)}} & \multirow{2}{*}{\makecell{\\[-0.8em]Sub-\mch}} & \multirow{2}{*}{\makecell{\\[-0.8em]Sub-\mch\ WD\\(C-O or He)}} & HV WD & $\cdots$                              & velocity shifts                    \\ \cmidrule(r){1-1}\cmidrule(l){5-7}
D$^6$ (two explosions)                  &                                &                                     &                                & None                          & $\cdots$                              & \makecell{velocity shifts\\+ double-peaked}                      \\
\colrule
WD collision                            & DET                            & Sub-\mch                            & Sub-\mch\ WD                   & None                          & $\cdots$                              & double-peaked                      \\
                                        &                                &                                     &                                & or HV 3$^\mathrm{rd}$ star\tnote{\,h}     &                                       &                                    \\
\colrule
\end{tabular}
\end{threeparttable}
\begin{tablenotes}
\item[a]{$^a$\,\textit{Delayed-DET}, delayed detonation; \textit{DEF}, pure deflagration; \textit{DET}, pure detonation.}
\item[b]{$^b$\,\textit{HV}, hypervelocity.}
\item[c]{$^c$\,We only report on peculiar spectroscopic signatures.}
\item[d]{$^d$\,A giant donor would survive the explosion partially stripped of its outer H/He-rich envelope, and be ejected from the system at a lower velocity compared to less massive companions.}
\item[e]{$^e$\,For weak deflagrations that fail to completely unbind the WD.}
\item[f]{$^f$\,Nebular line shifts are expected in an explosion ignited off-center.}
\item[g]{$^g$\,Such slow mergers can lead to an accretion-induced collapse to a neutron star and hence no \snia\ event.}
\item[h]{$^h$\,For triple systems consisting of a double-WD binary and a third low-mass star.}
\end{tablenotes}
\end{table}


\subsection{The classical Chandrasekhar-mass model, or single-degenerate (SD) scenario}\label{sect:mch}

Long considered the standard model for \sneia, this model involves a
WD that accretes material from a non-degenerate binary companion until
its mass approaches \mch. Here, the combustion must first proceed as a
deflagration and then transition to a detonation via the DDT mechanism
\citep{Khokhlov:1991}. Indeed, pure deflagrations result in weak
explosions that sometimes fail to completely unbind the WD (such
models have been proposed for Type Iax events), while pure detonations
are ruled out as they burn the entire WD to IGEs, with too little IMEs
needed to reproduce the early-time spectra\footnote{Pure detonations
of sub-\mch\ WDs however do produce the correct mix of IGEs and IMEs
since their initial density is lower compared to \mch\ WDs.}. Various
flavors of this delayed-detonation model exist, such as
gravitationally-confined detonations or pulsationally-reversed
detonations. The key requirement is that the WD is pre-expanded during
the deflagration phase before the detonation burns the remainder of
the WD at lower densities, producing IMEs in sufficient amounts.

The advantages of the SD scenario is that it naturally provides the
correct conditions to ignite carbon fusion in the WD core, in addition
to a physical basis (namely the \mch\ limit) for the observed
homogeneity of a large fraction of the \snia\ sample. Some
nucleosynthesis arguments (such as the abundance of manganese in the
solar neighborhood) point towards the need for a significant
contribution of \mch\ progenitors to the \snia\ population.
Chandrasekhar-mass delayed-detonation models have also been successful
at reproducing the observed light curves and spectra of normal \sneia,
although the range in \nifs\ yield needed to reproduce the observed
range in peak luminosity requires a somewhat artificial tuning of the
density at which the DDT occurs.

But this long favored model has several essential shortcomings.
Starting with stellar evolution, growing a WD via accretion to
\mch\ requires a fine tuning of the accretion rate to values of the
order of $10^{-7}$\,\msun\,yr$^{-1}$
\citep{Nomoto/Thielemann/Yokoi:1984}, and the predicted rates for the
SD channel are too low by at least one order of magnitude compared to
the observed \snia\ rate --- although these results are subject to
uncertainties related for instance to the common envelope (CE) phase
of the evolution, critical to forming close interacting binary
systems.  The SD scenario is also difficult to reconcile with the long
delay times of a few Gyrs between WD formation and explosion as
inferred from delay-time distributions (DTD; see
\citealt{Maoz/etal:2014}).  Last, some nucleosynthesis constraints
instead impose an upper limit on the fraction of \sneia\ resulting
from \mch\ progenitors \citep[e.g.][]{Bravo/etal:2022}.

The lack of a direct and unambiguous detection of the companion, even
in the favorable case of the nearby SN~2011fe
\citep[e.g.][]{Tucker/Shappee:2024}, as well as in \snia\ remnants in
our galaxy or in the Large Magellanic Cloud
\citep[e.g.][]{Kerzendorf/etal:2018}, is problematic. Indirect
detections include interaction of the rapidly-expanding ejecta with
the companion\footnote{Except in the ``spin up-spin down'' scenario,
where the WD must first rid itself of the excess angular momentum
gained via accretion before exploding, during which time the companion
continues to shed its envelope, shrinking considerably in size
\citep{Justham:2011}.}, leading to an early-time signature at UV-blue
wavelengths \citep{Kasen:2010}. Such signatures have been seen in
several \sneia\ but none have been unambiguously associated with
interaction with a companion. Likewise, H-rich material stripped from
the companion as it is overtaken by the SN ejecta should result in
narrow hydrogen lines in the nebular spectra, with only one firm
detection so far \citep{Prieto/etal:2020}.  Pre-SN mass loss via winds
from the companion, in addition to left-over material from the
accretion phase, should lead to a complex CSM with which the SN ejecta
interacts, resulting in radio signatures only detected in one event so
far (SN~2020eyj; \citealt{Kool/etal:2023}). This event also displayed
spectroscopic signatures of interaction with a He-rich CSM from a
helium star companion.  An X-ray signature is also expected in the SD
scenario, either during the pre-explosion phase as a supersoft X-ray
source, or resulting from CSM interaction. This has only been detected
in one event at nebular times \citep{Bochenek/etal:2018}, and no
detection has been reported even in the case of Ia-CSM events
\citep{Dwarkadas:2024}.  The pre-SN wind is also expected to carve out
a low-density cavity in the surrounding CSM, which has not been
convincingly detected in \snia\ remnants; these instead appear to be
expanding into a uniform medium \citep{Badenes/etal:2007}.


\subsection{Double-detonations in sub-Chandrasekhar-mass WDs with stable
  mass transfer}\label{sect:ddet}

Pure detonations of sub-\mch\ WDs provide a viable alternative to the
aforementioned \mch\ SD scenario. One immediate advantage is the
abundance of lower-mass WDs which could account for the observed
\snia\ rate. Second, the range of typical WD masses considered in such
models (0.8--1.1\,\msun) results in a range of density profiles and
hence a range in \nifs\ yields which naturally explains the observed
diversity in peak luminosity. In particular, the rapidly-evolving
light curves of low-luminosity 91bg-like \sneia\ seemingly require a
sub-\mch\ ejecta \citep{Blondin/etal:2017}.

In order to explode, sub-\mch\ WDs require the formation of a helium
shell around the C-O core, via accretion of H-rich material (which
burns to helium on the WD surface) or He-rich material from a donor
star.  Here we consider cases where the mass transfer is stable. This
helium shell must somehow detonate near its base, driving a powerful
shock wave into the WD, leading to a secondary detonation of the C-O
core. Initial models considered a relatively massive helium shell
($\lesssim 0.2$\,\msun), whose detonation produces copious amounts of
\nifs\ and other IGEs that are in conflict with the early-time
spectra. Subsequent theoretical investigations showed that less
massive helium shells (a few 0.01\,\msun) were able to detonate
and produce a successful secondary explosion while leaving little
IGE-rich material in the outer ejecta \citep[e.g.][]{Fink/etal:2010}.

There is a common misconception that the explosion of sub-\mch\ WDs
produces no stable isotopes of nickel. Despite the lower densities
involved in the burning ($\lesssim 10^{11}$\,\kgcm), stable \nife\ is
produced in NSE in sub-\mch\ explosions, albeit at a lower abundance
compared to \mch\ models at a given metallicity. The strength of the
nebular forbidden lines of nickel is highly sensitive to the
ionization state of the Ni-rich layers, such that abundance
determinations are uncertain \citep{Blondin/etal:2022}. Moreover,
there is no unambiguous correspondence between the stable nickel mass
and the mass of the exploding WD, since some \mch\ models
(e.g., gravitationally-confined detonations) predict similar stable IGE
yields as sub-\mch\ models. Nonetheless, \cite{Floers/etal:2020}
attempted to estimate the stable Ni/Fe ratio from late-time spectra,
and concluded that the majority of the \sneia\ in their sample were not
only consistent with sub-\mch\ explosions, but also inconsistent with
the predictions from standard \mch\ DDT models.


\subsection{Slow mergers or the classical double-degenerate (DD) scenario}

In this scenario, two unequal-mass sub-\mch\ C-O WDs in a close binary
system (whose total mass exceeds \mch) slowly spiral in via emission
of gravitational waves \citep{Webbink:1984,Iben/Tutukov:1984}. The
less massive (secondary) WD is disrupted and forms a hot envelope
surrounded by an accretion disk around the more massive (primary) WD.
The rapid accretion of C-O-rich material onto the primary WD ensures
an efficient increase in mass (possibly approaching \mch), at which
point the conditions for carbon ignition are reached in the core ---
as in the classical \mch\ scenario. The explosion will thus occur
within a massive C-O envelope, resulting in an interaction at early
times that could explain some super-luminous
\sneia\ \citep[e.g.][]{Taubenberger/etal:2013}.  However, the
spectroscopic properties of such an interaction have not yet been
thoroughly investigated.

This scenario naturally explains the absence of hydrogen and helium
lines in the spectra of \sneia, as well as the lack of any firm
detection of a non-degenerate companion star\footnote{This statement
also applies to all the other double-WD scenarios discussed in later
sections.}. Moreover, the loss of angular momentum via
gravitational-wave radiation results in a power-law $\mathrm{DTD}
\propto t^{-1}$ consistent with observations
\citep{Maoz/Mannucci:2012}.  The predicted rates of DD systems are
also closer to the observed \snia\ rate, although it has been argued
that mergers of sub-\mch\ WDs whose combined mass is less than
\mch\ must also contribute \citep{vanKerkwijk/etal:2010}. However,
surveys devoted to the discovery of close double-WD systems have only
discovered a handful of candidates that will merge within a Hubble
time, and only three systems with a combined mass either consistent
with \mch\ \citep{Geier/etal:2007} or slightly exceeding
\mch\ \citep{Pelisoli/etal:2021,Munday/etal:2025}.

One major drawback of the DD scenario is a result of the predicted
high accretion rate onto the primary WD. It is thought to lead to an
off-center ignition of carbon and the formation of a high-mass O-Ne
WD, with a subsequent accretion-induced collapse to a neutron star and
hence no \snia\ event \citep[e.g.][]{Shen/etal:2012}.


\subsection{Violent mergers}\label{sect:vm}

In contrast to the slow WD mergers discussed above, carbon ignition in
violent WD mergers occurs in the immediate precursor phase or during
the merger event itself. The first simulations showed that merging two
near-equal-mass $\sim 0.9$\,\msun\ WDs would lead to violent accretion
of the less massive WD onto the primary, and a hotspot would form at
the interface between the direct accretion stream and the C-O core of
the more massive WD, possibly resulting in carbon ignition and a
thermonuclear runaway \citep{Pakmor/etal:2010}. Since the exploding WD
is significantly sub-\mch, the amount of \nifs\ synthesized in the
explosion is rather small and only consistent with low-luminosity
\sneia. However, the larger ejecta mass results in longer diffusion
timescales, hence the predicted light curves are broader than for
classical 91bg-like events, although possibly compatible with peculiar
low-luminosity events such as SN~2010lp \citep{Kromer/etal:2013b}. The
secondary WD is only partially burned shortly after the explosion, and
its oxygen-dominated ejecta expands inside that of the primary WD,
which could explain the presence of narrow [\ion{O}{1}] lines in
nebular spectra of these sub-luminous events \citep{Kromer/etal:2016}.

Later studies considered unequal-mass WD systems with a more massive
primary WD ($\sim 1.1$\,\msun), resulting in \nifs\ yields compatible
with a normal \snia\ event \citep{Pakmor/etal:2012}. As in the
low-luminosity case above, the inner ejecta is dominated by the ashes
of the partially-burnt secondary WD, which consist mostly of oxygen
but also some neon. A strong forbidden line in the mid-IR range due to
[\ion{Ne}{2}]\,12.8\,\micron\ was indeed predicted to emerge in nebular
spectra of this violent merger model \citep{Blondin/etal:2023}, which
was detected soon after and for the first time in a JWST spectrum of
the Type Ia SN~2022pul (\citealt{Kwok/etal:2024}; see
Fig.~\ref{fig:nebspec}).

While a promising scenario for \sneia, it is unclear whether the
hotspot resulting from the direct-impact accretion stream actually
leads to a detonation, although longer-term spiral instabilities may
provide the necessary ignition conditions
\citep{Kashyap/etal:2015}. Also, the predicted level of asymmetry in
the ejecta can be fairly high, in contradiction with the low
polarization levels commonly observed in
\sneia\ \citep{Bulla/etal:2016} and with the spherical morphology of
\snia\ remnants.


\subsection{Double detonations in double-WD systems with dynamically unstable
  mass transfer (D$^6$ scenario)}\label{sect:d6}

The initial configuration for this explosion scenario is similar to
that for the violent mergers discussed above, and both result from a
dynamically unstable mass transfer. However, the explosion here is
triggered by the detonation of a thin helium layer on the primary WD
(either present initially or accreted from the secondary WD) --- as
opposed to direct carbon ignition in its core. This initial detonation
then leads to a secondary detonation of the C-O core, as in the
aforementioned classical double-detonation scenario. Such
explosions are referred to as helium-ignited violent mergers
\citep{Pakmor/etal:2013} or more poetically as dynamically driven
double degenerate double detonations (D$^6$;
\citealt{Shen/etal:2018}).

One important observational consequence compared to violent mergers is
that the secondary WD can survive the explosion and be ejected from
the system as a hypervelocity (HV) WD with peculiar chemical
composition. Several HV WDs have in fact been detected by the
\textit{Gaia} satellite, confirming the viability of this scenario,
although in too low numbers to explain the majority of
\sneia\ \citep{Shen/etal:2018b}.  However, if a thin helium shell is
also present on the secondary WD, the impact of the rapidly-expanding
ejecta from the primary can lead to its detonation and a subsequent
double-detonation explosion of the secondary, i.e., a
quadruple-detonation event \citep{Pakmor/etal:2022}. The predicted
delay of a few seconds between both explosions leads to the expansion
of the secondary WD's ejecta inside that of the primary, leading to
possible late-time signatures of asymmetries and peculiar chemical
profiles that could help distinguish quadruple detonations from cases
where the secondary WD survives the explosion.


\subsection{Collisions}\label{sect:coll}

There is a non-negligible probability for two WDs in dense stellar
environments (such as globular clusters or galactic nuclei) to collide
head-on with one another \citep{Benz/etal:1989}. Such collisions are
however expected to be rare, even when taking into account Kozai-Lidov
resonances induced by a third low-mass star orbiting the inner WD
binary \citep[e.g.][]{Katz/Dong:2012}.  Numerical simulations of
head-on collisions of WD pairs with varying masses and impact
parameters have been able to produce successful explosions with
\nifs\ yields that match the observed range in peak luminosity. A
bimodal \nifs\ distribution is often expected in such models,
resulting in double-peaked emission lines of cobalt and iron at late
times that have been detected in a few nebular \snia\ spectra
\citep{Dong/etal:2015}. It is still unclear however whether these
features are a smoking-gun signature for this scenario, as
double-peaked profiles could also result from quadruple detonations in
the D$^6$ scenario discussed above.


\section{Ongoing and future developments}\label{sect:future}

Here we mention a few ongoing and future efforts to better understand
the nature of \sneia, both from observational and theoretical/modeling
perspectives.


\subsection{Observations}

There already exists a plethora of high-quality photometric and
spectroscopic data for \sneia\ collected over several decades.  That
said, more recent high-cadence surveys --- such as the Asteroid
Terrestrial-impact Last Alert System (ATLAS;
\citealt{Tonry/etal:2018}) or the Zwicky Transient Facility (ZTF;
\citealt{Bellm/etal:2019}) --- have enabled the systematic discovery of
\sneia\ within a few days from the explosion. Studies of the earliest
phases of the evolution place important constraints on the presence
of a (massive) binary companion or CSM, or on the presence of
\nifs\ in the outer ejecta.

The upcoming Vera Rubin Observatory, with its 10-year Legacy Survey
of Space and Time (LSST), will discover and monitor over a million
\sneia\ starting in 2025. The observing cadence, however, will result
in relatively sparsely-sampled light curves, and spectroscopic
follow-up is not included in the survey design and will be limited to
a small fraction of the total sample. Rubin/LSST will also discover
over a hundred \sneia\ out to redshifts $z \lesssim 1$
that are strongly lensed by a foreground galaxy
\citep{Arendse/etal:2024}, producing multiple images of the SN that
appear at different times. By detecting the first appearing SN image,
one can use the predicted time delays between the multiple images to
catch the SN right at the moment of its next appearance, and hence
access the first hours/days of its evolution
\citep{Suyu/etal:2020}. Moreover, the rest-frame UV radiation of these
high-redshift events can be observed with ground-based facilities at
optical wavelengths, providing complementary diagnostics to
high-cadence surveys targeting nearby events.

The recent launch of JWST has enabled the systematic study of
\sneia\ at mid-IR wavelengths --- such data only existed for a handful
of events before then.  The first late-time spectra of
\sneia\ obtained with JWST revealed for the first time the
2.5--5\,\micron\ range for these events, as well as the detection of
the [\ion{Ne}{2}]\,12.8\,\micron\ line in one event, as predicted by
violent merger models (see Fig.~\ref{fig:nebspec}). The
\textit{Euclid} satellite, although not designed for transient
surveys, could be used to complement the high-redshift \snia\ sample
of Rubin/LSST for cosmological studies at near-IR wavelengths
\citep{Bailey/etal:2023}. The future Nancy Grace Roman Space Telescope
(NGRST), on the other hand, will carry out a dedicated near-IR survey
of a few thousand
\sneia\ \citep{Hounsell/etal:2018}\footnote{\cite{Hounsell/etal:2018}
refer to the WFIRST project, since then rebaptized NGRST.}.

At higher frequencies, the X-ray satellite XRISM (launched in
September 2023), and possibly succeeded by ESA's NewAthena mission in
the late 2030s, will carry out spatially-resolved spectroscopic
studies of \snia\ remnants, thereby directly constraining the
explosion energetics and ejecta chemical stratification.  In the
$\gamma$-ray regime, the Compton Spectrometer and Imager (COSI;
planned launch in 2027) should be able to detect energetic photons
resulting from the $^{44}$Ti decay chain, whose abundance can be used
to discriminate between different \snia\ explosion mechanisms
\citep[e.g.][]{Kosakowski/etal:2023}.

Last, the revolution wrought by the direct detection of gravitational
waves has opened up a new channel for studying astronomical events,
including \sneia. Double WD systems are expected to largely dominate
over all other compact binaries detected by LISA in our galaxy
\citep{Amaro-Seoane/etal:2023}, with an expected $\gtrsim$2000 WD+WD
systems with a primary mass exceeding 0.8\,\msun\ that could explode as
\sneia\ (\citealt{Korol/etal:2024} and private communication). LISA
will also be able to detect WD+black hole binaries and test the
scenario in which a WD is tidally stripped by an intermediate-mass ($<
10^5$\,\msun) black hole \citep{Rosswog/etal:2009a}.


\subsection{Theory and Modeling}

On the theoretical and modeling fronts, ongoing efforts are focused on
stellar evolution, the hydrodynamics of the explosion and its coupling
to the nucleosynthesis, and the radiative-transfer post-processing.

Concerning stellar evolution, 3D modeling of the common envelope phase
of evolution is currently being undertaken to better understand how
close binary systems can form.  More accurate rate estimates for the
\mch\ channel require a better understanding of accretion processes
onto WDs that can result in a near-\mch\ progenitor, as well as of the
pre-explosion simmering phase.

Numerical modeling of the explosion still suffers from insufficient
resolution to properly model the combustion fronts in \sneia, although
recent simulations have been able to resolve the detonation of a thin
helium layer atop a C-O WD, with a grid-cell resolution of a few tens
of meters only (F.~R\"opke, private communication). However,
nucleosynthesis calculations still largely rely on post processing
explosion simulations via tracer particles, where one would prefer a
direct coupling of nuclear reaction networks to the hydrodynamics
\citep{Bravo:2020}. The predicted abundances of various isotopes are
however only weakly sensitive to uncertainties in nuclear reaction
rates \citep{Bravo/Martinez-Pinedo:2012}. Last, terrestrial combustion
experiments, in particular related to the DDT mechanism in unconfined
media, can provide useful insights into the treatment of combustion
fronts in simulations of the explosion \citep{Thomas/etal:2022}.

Radiative-transfer simulations still represent a major hurdle in
connecting explosion models with observations (light curves and
spectra). While a 3D treatment of the problem seems warranted, there
is a trade-off to be made between the dimensionality and physical
consistency or resolution of the simulation, and full time-dependent
non-LTE 3D simulations are currently lacking. One related major
limitation concerns the availability of atomic data, whose accuracy
can have a significant impact on spectroscopically determined
abundances \citep{Blondin/etal:2023}. Sophisticated numerical
calculations are required to compute cross-sections for important
atomic processes (e.g., photoionization, collisional excitation) as
these cannot all be measured in the laboratory.


\section{Concluding remarks}\label{sect:ccl}

Much progress has been made in recent years towards a better
understanding of the progenitors and explosion mechanisms of
\sneia. High-cadence surveys have enabled to systematically discover
new \sneia\ shortly after explosion and hence enable the study of
their earliest phases, and JWST has given the community access to
high-quality spectroscopic diagnostics in the mid-IR range. While
future observational efforts will concentrate on gathering large
samples of \sneia\ in the optical and near-IR ranges for cosmological
studies, detailed studies of individual events (including radio, X-ray
and $\gamma$-ray observations when possible) remain paramount for
constraining key parameters of the explosion.

For a long time the single most-favored contender for explaining the
bulk of \sneia, the \mch\ model is now invoked to explain more
peculiar events, such as Type Iax SNe which are thought to result from
pure deflagrations of \mch\ WDs. The recent emergence of double
sub-\mch\ WD systems as the leading model for \sneia\ is an
interesting reversal of the general consensus up to a decade ago, and
testifies to the dynamism of this field of research.  However, a word
of caution is in order, as future observational findings and modeling
efforts might be able to address some of the limitations of the
\mch\ model. Likewise, future gravitational-wave detections of
double-WD systems with LISA will be able to confirm or disprove the
double-degenerate scenario as the dominant channel for \sneia. The
jury is still out, and new ideas on how white dwarf stars explode will
no doubt emerge in the coming years.


\begin{ack}[Acknowledgments]

This chapter is dedicated to Tom Marsh (1961--2022), who introduced SB
to the topic of Type Ia supernovae during his studies at the
University of Southampton.
We thank Eduardo Bravo, Robert Fisher, Bruno Leibundgut, and Ken Shen
for carefully reading and providing detailed comments on a first
draft of this chapter.
We further acknowledge useful discussions with Luc Dessart, Roland
Diehl, Valeriya Korol, Friedrich (Fritz) R\"opke, Ivo Seitenzahl, with
a special mention to Frank Timmes for inspiring the ``Three basic
ingredients for a successful \snia\ model'' box.
We thank Stefano Benetti for providing the light-curve data for
SN~2002ic used in Fig.~\ref{fig:pec} and Eugene Churazov for providing
the SN~2014J data and model shown in Fig.~\ref{fig:co56}.
This work was supported by the Alexander von Humboldt foundation, and
by the `Action Thématique de Physique Stellaire' (ATPS) of CNRS/INSU PN Astro
co-funded by CEA and CNES.
\end{ack}

\seealso{Several chapters of the ``Handbook of Supernovae''
(Eds. A.~W.~Alsabti and P.~Murdin), four of which are referenced in the
present chapter, cover many aspects of
\sneia\ in more depth, including their cosmological applications.
Many explosion models for \sneia\ are publicly available online at the
Heidelberg Supernova Model Archive (HESMA;
\url{https://hesma.h-its.org}).
The results of radiative-transfer simulations for supernova
ejecta with the CMFGEN code are accessible on Zenodo
(\url{https://zenodo.org/communities/snrt}), and the results of a
radiative-transfer code-comparison study are available on github
(\url{https://github.com/sn-rad-trans/data1}; see
\citealt{Blondin/etal:2022b}).
}

\bibliographystyle{Harvard}
\bibliography{snia_blondin}

\end{document}